\documentclass[11pt,a4paper,oneside]{article}
\usepackage[top=3cm, bottom=3cm, left=2cm, right=2cm]{geometry}
\usepackage[english]{babel}
\usepackage[utf8x]{inputenc}
\usepackage[T1]{fontenc}

\usepackage{amsthm,amsmath,amssymb,mathrsfs,dsfont,mathtools}

\usepackage{bm}
\usepackage{bbm}
\usepackage{graphicx}
\usepackage[colorinlistoftodos]{todonotes}
\usepackage[colorlinks=true, allcolors=blue]{hyperref}
\usepackage{ulem}
\usepackage{booktabs,subcaption}
\usepackage[all]{nowidow}
\usepackage{setspace}
\usepackage{algpseudocode}
\usepackage{algorithm}
\usepackage{tikz} 
\usepackage[sf]{titlesec}
\usepackage{sectsty}
\allsectionsfont{\centering \normalfont\scshape} 
\usepackage{lipsum}
\usepackage{adjustbox}
\usepackage[running]{lineno} 
\usepackage[round]{natbib}

\usepackage[all]{nowidow}
\usepackage{setspace}
\usepackage{algpseudocode}
\usepackage{algorithm}
\usetikzlibrary{decorations.pathreplacing,arrows.meta, positioning,chains,fit,shapes,calc}

\usepackage{flexisym}
                   
\algnewcommand{\Inputs}[1]{%
  \State \textbf{Inputs:}
  \Statex \hspace*{\algorithmicindent}\parbox[t]{.8\linewidth}{\raggedright #1}
}
\algnewcommand{\Initialize}[1]{%
  \State \textbf{Initialize:}
  \Statex \hspace*{\algorithmicindent}\parbox[t]{.8\linewidth}{\raggedright #1}
}

\usepackage{authblk}
\title{Link prediction in ecological networks under extreme taxonomic bias}
\date{}
\author[1]{Jennifer N. Kampe}
\author[2]{Camille M.M. DeSisto} 
\author[1]{David B. Dunson}

\affil[1]{Department of Statistical Science, Duke University, Durham, NC, 27708, U.S.A.}
\affil[2]{Nicholas School of the Environment, Duke University, Durham, NC, 27708, U.S.A.}

\theoremstyle{definition}

\newcommand{\bbeta}{\mathbf{\beta}}
\newcommand{\bgamma}{\mathbf{\gamma}}
\newcommand{\bdelta}{\mathbf{\delta}}
\newcommand{\bzeta}{\mathbf{\zeta}}

\newcommand{\bA}{\mathbf{A}}
\newcommand{\bF}{\mathbf{F}}
\newcommand{\bL}{\mathbf{L}}
\newcommand{\bO}{\mathbf{O}}
\newcommand{\bC}{\mathbf{C}}
\newcommand{\bU}{\mathbf{U}}
\newcommand{\bV}{\mathbf{V}}
\newcommand{\bW}{\mathbf{W}}

\newcommand{\bX}{\mathbf{X}}
\newcommand{\bP}{\mathbf{P}}

\newcommand{\one}{\mathbbm{1}}
\newcommand{\Nf} {\mathcal{N} }

\begin{document}

\maketitle
\textbf{Corresponding author}: Jennifer N. Kampe, Department of Statistical Science, 404A Old Chemistry, Box 90251, Durham, North Carolina, 27708, U.S.A. \textbf{Email}: jennifer.kampe@duke.edu\\

\newpage

\begin{abstract}
1. Ecological networks offer powerful insights into community function but, without first characterizing these networks accurately, our ability to detect and interpret changes under environmental stress is limited. \\
2.  We develop an approach to reduce bias in link prediction in the common scenario in which data are derived from studies focused on a small number of species. Our Extended \textbf{Co}variate-\textbf{I}nformed \textbf{L}ink Prediction (COIL+) framework employs a latent factor model that flexibly borrows information across species, incorporates species traits and phylogeny, and leverages information from multiple studies to address uncertainty in species occurrence. We also propose a trait-matching procedure that allows heterogeneity in species-level trait–interaction associations. We illustrate the approach with a literature-based dataset of 268 sources reporting Afrotropical frugivory and compare performance with and without correction for occurrence uncertainty. \\
3. COIL+ substantially improves link prediction and reduces sampling bias, revealing 5,637 likely but unobserved frugivory interactions (a median of nine additional interactions per frugivore). Newly predicted interactions are concentrated among poorly sampled frugivores, such as the water chevrotain (\textit{Hyemoschus aquaticus}, a small forest-dwelling ungulate) and the rufous-bellied helmetshrike (\textit{Prionops rufiventris}, a passerine bird of East African tropical forests). Additionally, the method improves model discrimination compared to existing methods under strong taxonomic bias and narrow study focus.\\
4.  This framework generalizes to diverse network contexts and provides a useful tool for link prediction in the face of biased interaction data.
\end{abstract}

\textbf{Keywords}: Bayesian methods; Interaction network; Latent factors; Link prediction; Taxonomic bias

\newpage

\section{Introduction}

Species interactions form the foundation of ecological communities: Their study is crucial to understanding both the natural processes influencing community assembly and the likely impacts of anthropogenic activities. The collection of biotic interactions within a community or ecosystem can be described as a network whose topology contains intrinsic interest and is also instrumental in the study of ecological resilience \citep{suweis_emergence_2013}. A growing body of literature suggests a powerful link between network structure and community dynamics, namely robustness in the face of disturbance \citep{Dunne}, tolerance to extinction \citep{Memmott}, and productivity \citep{Fiegna, Duffy}. As a result, a more complete understanding of network structure increases the ability of researchers to accurately predict the impacts of disruptive processes, including habitat destruction, pollution, and climate change. Moreover, monitoring of interactions is of scientific interest, as the loss of an established pattern of interaction is the loss of an ecological function \citep{Jordano}.

The study of these networks poses a persistent challenge to ecologists due to the complexity of obtaining and analyzing network data. Sampling interactions requires extensive fieldwork, often substantially more intense than that required for biodiversity surveys \citep{Jordano}. Different interaction data collection methods are prone to varying logistical feasibility and financial resources; for example, direct observations tend to require challenging logistics but costs may be lower, while DNA-based molecular techniques are associated with high financial burdens \citep{quintero_methodological_2022}. Many ecological interactions remain unobserved by researchers, even when resources allow a high sampling effort \citep{Chacoff2012}. Most ecological networks are therefore under-sampled. Compiling data from many different studies into meta-networks (e.g., \cite{durand-bessart_trait_2023, fricke_accelerating_2020, desisto_drivers_2022}) can help reduce under-sampling, advancing the study of global ecology and biogeography \citep{windsor_using_2023}. Nevertheless, meta-networks are typically characterized by high sampling bias (e.g., \cite{tonos_examining_2024}).

Network studies of plant-animal interactions such as frugivory, pollination, and herbivory can broadly be categorized as either \textit{zoocentric}, \textit{phytocentric}, or combined. In the zoocentric (animal-centered) approach, interactions are observed via animal activity, including direct behavioral observations and camera traps. Researchers also use microscopic, metabarcoding and stable isotope analysis of fecal samples to identify frugivory interactions \citep{Gonzalez, quintero_methodological_2022} and pollen samples recovered from pollinator bodies to identify pollination interactions \citep{Olesen}. Zoocentric studies improve detection of interactions involving rare plants, yet tend to undersample animal species that do not forage in the understory \citep{Vitorino}. The phytocentric approach, in contrast, often favors more common plants; direct observation of feeding events can be obtained by monitoring focal species \citep{Cirtwill}, sampling plots, or surveying transects containing focal species. Alternatively, focal plant species can be sampled and inspected for herbivory marks, with associated eDNA analyzed \citep{Kudoh}. Combined‐focus studies may use a mixture of techniques (e.g., field observation, literature surveys, and ethnobotanical interviews) often resulting in more complete networks \citep{Jordano, quintero_methodological_2022}. In particular, surveys leveraging local ecological knowledge provide valuable interaction records in logistically challenging or hyper-diverse systems \citep{ong2021building, desisto2025structure}. Although studies employing a combination of methods tend to generate a greater number of links \citep{Escribano}, undersampling remains a significant concern \citep{Jordano}.

Incompleteness of sampled networks is exacerbated by study-specific taxonomic and geographic bias \citep{PapadogeorgouJASA}. Network studies are necessarily limited to a specific geographical area and can only record interactions for the subset of species whose ranges lie within the study area. An interaction that occurs in one region may not appear in a second area where the relevant species co-occur, due to spatial variability in species abundance, phenology, and the presence or absence of preferred food sources \citep{Poisot}. Simultaneously, limitations imposed by the research questions of interest typically limit network studies to a small number of focal species. As a result, while a meta-network assembled from multiple network studies may reduce under sampling on the whole, it will be biased by asymmetry of sampling effort across taxa and locations \citep{tonos_examining_2024}. 

A further, fully insurmountable source of bias for observational studies is that posed by \textit{forbidden interactions}, i.e., those which are prevented by non-overlapping geographical ranges, non-overlapping phenologies, or incompatible morphologies \citep{Vitorino}.  Any network sampled through field work is intrinsically incapable of detecting forbidden interactions, such as pollination of a flowering plant by an insect not present in its range. However, these potential interactions may be of interest to community ecologists looking to predict the likely impacts of shifting ranges and phenologies that are increasingly observed due to climate change, or to epidemiologists seeking to predict pathogens with a high potential to jump from non-human primates to humans \citep{Werner}.

Pervasive under-sampling and bias in observed networks have motivated a rich and growing statistical literature on link prediction, with particular attention paid to dimension reduction. The collection of biotic interactions within an ecological community can be efficiently represented as a network in which species are represented by nodes, and interactions are represented by edges between the nodes. In most practical scenarios, network researchers observe presence-only data for a relatively small number of edges and are interested in performing inference on all possible edges. In practice, for a network representing interactions among $n_F$ frugivores and $n_P$ plants, it is of interest to predict whether or not an interaction exists among all $n_F n_P$ possible frugivore-plant pairs, while only a relatively small number of interactions ($n$) have been observed. This creates an \textit{ill posed} statistical problem in which the number of parameters ($n_F n_P$) vastly exceed the number of observations ($n$); such a problem may not have a unique solution and hence inferences from the model may be invalid.

Latent space representations offer a flexible class of solutions, providing efficient inference and prediction on the basis of biased and under-sampled networks. Pioneered by \cite{Hoff}, the latent space approach to network analysis permits massive dimension reduction: instead of estimating a separate parameter for all species pairs, such models embed each species in a low-dimensional Euclidean space and estimate the interaction propensity between any given pair via their distance in the latent space. Recent advances in ecological network modeling incorporate domain-specific knowledge into network models, including phylogenetic and morphological information. A number of recent studies point to the importance of morphological traits in driving dietary interactions \citep{Dehling, Goldel}. It is not surprising then that including traits in interaction models has proven fruitful: \cite{Cardoso} and \cite{dallas2019host} find that mammalian hosts with similar functional traits share parasite species. Where morphological data are limited, the phylogenetic or taxonomic relationship can play an important role in explaining interaction patterns between species \citep{Werner}, as key traits can be linked to conserved genetic regions. 

Recent work by \cite{PapadogeorgouJASA} synthesizes cutting-edge ecological approaches within the classical latent space network model. This model introduces two innovations which are highly valuable in the context of bipartite species interaction networks -- particularly meta-networks. First, it links species-level traits and phylogenies to latent factors, allowing observable traits and unobservable species detectability to inform the latent embedding. Second, it uses a conditional likelihood specification to account for the differential effort caused by taxonomic and geographic biases in the study design. This results in a rich model which is highly successful at describing patterns of connectedness in meta networks. However, in practice, the success of this model is highly dependent on the availability of network-wide studies. 

A prerequisite for bias correction network methods such as \cite{PapadogeorgouJASA} is data describing which interactions are detectable under the respective study's methods, particularly study focus and species co-occurrence patterns. Although study focus can be inferred with a high degree of confidence, regional occurrence and co-occurrence information is not often available in single-species studies or studies with \textit{extreme taxonomic bias}.

While nearly all observational network studies result in presence-only data, multi-species studies provide some information about non-interactions as well as interactions. When a study targets multiple focal species, it provides more comprehensive link detection, but also a more complete picture of local species occurrence. In such a study, non-interactions can be informative: if a zoocentric study involves two focal monkey species within the same region, and only one of them is observed eating the fruit of a given plant, this provides some moderate evidence that the other monkey species does not feed upon that plant. Because we know that the plant and both monkey species indeed co-occur, the absence of an observed interaction suggests the lack of a true interaction. If, in contrast, our study is zoocentric or phytocentric with a single focal species, the informativeness of non-edges is highly uncertain: absence may be due either to a lack of co-occurrence or lack of a true interaction. 

Single-species studies hence provide very little information about non-interactions, introducing additional difficulty into the challenge of link prediction. In the Model Details in Section \ref{ssec:model} below, we propose a solution that incorporates additional uncertainty in species occurrence to improve the informativeness of non-edges in data sets dominated by studies with extreme taxonomic bias, i.e., either zoocentric or phytocentric studies with a small number of focal species. Although the focus of our study is on plant-animal interactions, specifically frugivory, the methods are relevant to other ecological network types such as host-pathogen, predator-prey, plant-mycorrhizal, and microbiome.

\section{Materials and methods}

We begin with a description of the Afrotropical frugivory dataset, consisting of a corpus of taxonomically and geographically biased interaction studies which provide the response variable for our network analysis, as well as supplemental data used to construct trait and phylogenetic explanatory variables. We then describe our proposed method for link prediction with an emphasis on the appropriate construction of study metadata variables and the incorporation of expert knowledge. 

\subsection{Data}

\subsubsection{Afrotropical Frugivory Database} \label{sec:AF_data}

We used observed interaction data from \cite{durand-bessart_trait_2023}, a compilation of literature that represents the most complete data available for frugivory interactions in Afrotropical forests. We update this database by conducting a literature search for Afrotropical frugivory interactions on Google Scholar, leading to the identification of 32 relevant articles that were not already included in the \cite{durand-bessart_trait_2023} database. The combined Afrotropical Frugivory dataset consists of 10,635 records spanning 5,938 unique pairwise frugivory interactions involving 267 frugivore species and 958 plant species. The data were extracted from 268 sources spanning at least 327 sites across Afrotropical forests. Of these, only 1\% are full network studies, while the vast majority (82\%) are zoocentric, 6\% are phytocentric, and 10\% focus on a single frugivore-plant pair.  Of the zoocentric sources, 66\% demonstrate \textit{extreme taxonomic bias}, focusing on a single frugivore. 

\begin{figure}[H]
    \centering
    \includegraphics[width = 0.85\textwidth]{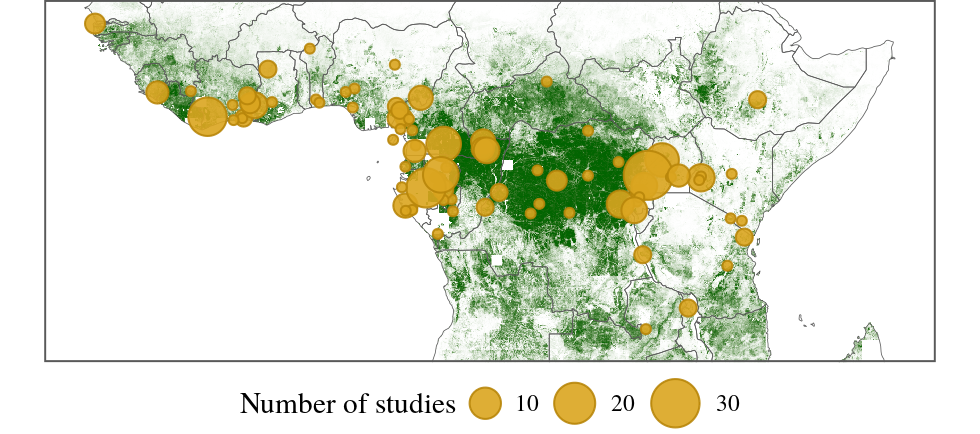}
    \label{fig:my_label}
    \caption{Study locations are distributed throughout Afrotropical forests with significant geographical bias favoring the eastern and western edges of the Congo river basin; green overlay indicates tree cover based on data provided by \cite{reiner_treecover}. The dataset includes 66 sources which do not provide an explicit location and are omitted from the map. }
\end{figure}

In the abstract, the Afrotropical Frugivory Database, and indeed any data suitable for the proposed method, is a meta-network that provides (1) a collection of observed interactions drawn from multiple different studies, and (2) information on the focal species for each study. Although the latter component of the data is not always explicitly provided within each study, the study type (e.g., zoocentric, phytocentric) is provided and allows recovery of focal species; this process is illustrated in the Model Details in Section \ref{ssec:model} below. 

\subsubsection{Supplemental Data} \label{sec:AF_supplemental data}

Because species that share a long portion of their evolutionary history are likely to display similar interaction propensities, phylogenetic correlation matrices provide an important source of data for learning interactions. Phylogenetic correlation matrices are computed via the \texttt{R} package \texttt{ape} \citep{ape_2019}. For frugivores, a consensus tree is obtained from VertLife \citep{vertlife}, while plant phylogenetic data are acquired using the \texttt{V.PhyloMaker} package \citep{V.phylo} in \texttt{R}, making use of the most up-to-date angiosperm phylogeny in \cite{plat_phylogeny} generated from GenBank and Open Tree of Life data.

Physical traits of frugivores and plants provide additional information on interaction propensity, a widely known phenomenon known as trait-matching. We include the following traits for frugivores, selected due to their widespread availability and relation to energy requirements and foraging habits: natural logarithm of body mass \citep{durand-bessart_trait_2023, iucn_iucn_2024, jonathan_kingdon_kingdon_2015}, generation length \citep{pacifici_generation_2013, iucn_iucn_2024}, conservation status based on the International Union for the Conservation of Nature (IUCN) Red List \citep{iucn_iucn_2024}, and habitat (forest or mixed) \citep{iucn_iucn_2024}. Plant trait data include wood density \citep{rejou-mechain_biomass_2017}, fruit width, and fruit length \citep{durand-bessart_trait_2023, DeSistoKampe}. We include both fruit length and width in models; while both variables are related to frugivory, the fruit width is more sensitive to frugivory-driven selection than the length of the fruit \citep{yu_contrasting_2024}. Additional details on the preparation of trait data are provided in Supplemental Materials.

\subsection{Model Details} \label{ssec:model} 

We begin by describing the structure of an appropriate dataset in the abstract, building on the Afrotropical Frugivory dataset introduced above for intuition. We next introduce the bias correction mechanism via conditional likelihood specification and then provide the details for the covariate-informed latent factor model underlying link prediction.

\subsubsection{Data Structures and the Conditional Likelihood}

Assume that interactions involving $n_F$ frugivore species and $n_P$ plant species are observed in $n_S$ studies. Interactions observed in each study are recorded in a binary adjacency matrix; the collection of adjacency matrices is stacked into a three-dimensional array $\bA$ with dimensions $n_F \times n_P \times n_S$ whose entries indicate the observation of an interaction: $A_{ijs} = 1$ if frugivore species $i$ and plant species $j$ were observed in an interaction in study $s$ and $A_{ijs} = 0$ otherwise. 

The motivating problem is that the observed adjacency array $\bA$ is incomplete due to under-sampling, taxonomic and geographical bias, and non co-occurrence (i.e., forbidden interactions). We are interested in recovering true interactions, defined as \textit{those interactions that would take place, given the opportunity}. Let $\bL$ be the $n_F \times n_P$ latent variable matrix of true interactions: we are interested in learning unobserved interactions, i.e., frugivore-plant pairs in the set $\{i,j: L_{ij} = 1, A_{ijs} = 0 \text{ for all } s = 1, \cdots, n_S\}$.

To impute these unobserved interactions, we must first define a probability model which reflects the core fact that interactions between non-co-occurring or non-focal species pairs are unobservable. In particular, the likelihood for any single interaction follows the conditional formulation of \cite{PapadogeorgouJASA}:
\begin{equation}\label{eq:Likelihood}
   P(A_{ijs} = 1 \mid L_{ij} = l, F_{ijs} = f, O_{ijs} = o)=\left\{
  \begin{array}{@{}ll@{}}
      0, & \text{if } lfo =0 \\
      p_i q_j, & \text{if } lfo =1
  \end{array}\right.
\end{equation} 
where $O_{ijs}$ and $F_{ijs}$ are co-occurrence and focus indicators for species $i$ and $j$ in study $s$, and $p_i, q_j$ are detection probabilities for species $i$ and $j$, respectively. Additionally, $(l,f,o) \in \{0,1\}^3$ are particular values of the respective binary indicators for the presence of a true interaction, study focus, and co-occurrence. Hence, an interaction is observable only if it involves focal, co-occurring species with a true interaction; in this case it is observed with a probability given by the product of the relevant species detection probabilities. Additional details on the key focus and occurrence data structures, as well as the learned species detection parameters, are provided below.

The study \textit{focus} refers to the taxonomical breadth of a study, or the set of bipartite interactions that would have been recorded if observed. The focal array $\bF$ has dimensions equal to that of the observed interaction array ($n_F \times n_P \times n_S$) and has entries $F_{ijs} = 1$ if an interaction between frugivore $i$ and plant $j$ would have been recorded in study $s$, had it been observed, and $F_{ijs} = 0$ otherwise. As in the case for the Afrotropical Frugivory database, study focus can be recovered given study type (e.g. zoocentric or phytocentric) and the observed adjacency matrix, as follows. If study $s$ is zoocentric, the focal frugivores are given by the set of all frugivore species $i$ such that at least one interaction is observed in study $s$: $F^F_s =\{i: \sum_j A_{ijs} > 0 \}$. Because the hypothetical study is zoocentric, any interaction involving a focal frugivore will be recorded, hence the focal array slice $F_{..s}$ has $F_{ijs} = 1$ for all $\{i,j: i \in F^F_s, j = 1,2. \cdots, n_P\}$. Similarly, for a phytocentric study $s$, we define $F^p_s$ to be the set of all focal plants: $F^P_s=\{j: \sum_i A_{ijs} > 0 \}$. For the corresponding entries of the focal  array slice $\bF_{..s}$, we have $F_{ijs} = 1$ for all $\{i,j: j \in F^P_s, i = 1,2. \cdots, n_F\}$. Finally, in a combined or ``network'' study $s$, $F_{ijs} = 1$ for all $i,j$, i.e., all observed interactions are recorded. Study type can be easily learned from the title or abstract of the study.  

The next data structure needed to construct the likelihood in Equation \ref{eq:Likelihood} is the co-occurrence array $\bO$ with dimensions  $n_F \times n_P \times n_S$ (the same as the interaction and focus arrays), and having entries $O_{ijs} = 1$ if frugivore species $i$ and plant species $j$ are both present in the site for study $s$ during the study period, and $O_{ijs} = 0$ otherwise. Because these data are not explicitly collected or provided in a typical network study, we assume that they can be estimated \textit{with uncertainty} as a latent data structure. Note, however, that the model easily accommodates known occurrence information in the rare event that it is provided; in the description below we assume that it is not. 

To illustrate the influence of occurrence, we first consider a naive approach which assumes that only species observed in a frugivory interaction in study $s$ are present during study $s$. Consider the following scenario, common in the Afrotropical Frugivory dataset: study eight, \cite{Tchamba}, is a zoocentric study focused on the dietary habits of the African forest elephant (\textit{Loxodonta cyclotis}) in the Santchou Reserve, Cameroon. Because study eight considers a single focal frugivore species, this study has extreme taxonomic bias, limiting our observation of plants to those that are fed on by the forest elephant. Using the naive occurrence construction above, we set $O_{ij8} = 0$ for all plants not observed to be fed on by \textit{Loxodonta cyclotis}. In this case, non-edges in $A_{ij8}$ will not contribute to the likelihood defined in Equation \ref{eq:Likelihood}, and hence non-edges are \textit{non-informative}. As a result, the model will tend to overestimate interaction prevalence for \textit{Loxodonta cyclotis}, having been trained on data which ostensibly shows that the forest elephant eats any plant it comes across. While, in this case, the deficiencies of our occurrence construction may be masked by the fact that elephants are generalists, they are known to avoid certain fruits, e.g., citrus. Hence, even in this most favorable case, link prediction will suffer from a high false positive rate.

If the above study instead followed multiple animal species, the naive occurrence construction provided above would be more reliable: assuming that different frugivore species have different dietary preferences, an increase in the number of focal frugivore species will increase the accuracy of the occurrence information implied by the observed adjacency matrices. However, in the absence of multi-species and network-wide studies, additional methods are needed to overcome the selection bias introduced by taxonomically narrow studies. 

To avoid the pitfalls of occurrence misspecification, we provide a method for incorporating uncertainty into occurrence estimates in the case of extreme taxonomic bias. If the occurrence information is definitively known due to concurrent abundance assays, then $O_{ijs}$ can be assumed to be known and specified directly as data. In the more typical case, $O_{ijs}$ is a latent variable imputed in the model. 

In the absence of independent occurrence data, the model takes as input a set of occurrence probabilities, assumed to be known \textit{with uncertainty}, separately for the two types of species. In our application, we construct a prior occurrence probability matrix for frugivores and plants: $\bP_{O^F}, \bP_{O^P}$ having dimensions $n_F \times n_S$ and $n_P \times n_S$ respectively, with entries $p_{O_{is}^F}$ ($p_{O_{js}^P}$) indicating the prior probability that species $i$ ($j$) was present in study $s$. These user-provided probabilities are used as the prior means for the uncertain occurrence probabilities, which will be learned by the model. This specification allows us to account for uncertainty in both occurrences and associated probabilities, reducing the sensitivity of the model to uncertain user inputs. 

The key question then is where the user-defined prior occurrence probabilities come from. Often data from abundance surveys may be available in the absence of network studies; indeed, such studies are typically less challenging and more commonly executed. Nonetheless, the most common scenario is that no independent surveys are available or that abundance surveys do not perfectly correspond to study areas. We propose a simple method of using expert knowledge to construct prior occurrence probabilities which borrow information across ecologically similar studies. Using the Afrotropical Frugivory data as an example, presence-only data for frugivores and plants in each study can be extracted from the respective row and column sums of each observed interaction adjacency matrix. Site meta-data including GPS coordinates, country, region, and habitat type can then be used to link sites, with ecological knowledge being used to specify the degree of information borrowing across sites, e.g. a species previously present in the same nature preserve may be given a prior occurrence probability of 75\% at subsequent studies at the same site, while a species only previously observed in the same region and habitat type might be given a much smaller probability of occurrence of 15\%. Because these probabilities provide the mean of a weakly informative prior on model-estimated occurrence probabilities, uncertainty is appropriately incorporated through model specification. In the absence of expert knowledge, a default prior occurrence probability for species which are not observed during a given study can be used; recommendations for default occurrence probabilities are provided in the Supplement. 

To illustrate the incorporation of uncertainty in occurrences, we briefly describe the updating of occurrence indicators within the model here, using frugivore occurrences for illustration. When $\sum_{j} A_{ijs} > 0$, some interaction involving frugivore species $i$ is observed in study $s$, so species $i$ is known to have occurred in study $s$; neither the occurrence probability nor the occurrence indicator is sampled in this case. If, in contrast $\sum_{j} A_{ijs} = 0$, occurrence for frugivore $i$ at site $s$ is unknown. In this case, the user-specified occurrence probabilities are used as the center for weakly informative truncated normal priors placed on occurrence probabilities:
$$\pi_{O_{is}^F} \sim \Nf(p_{O_{is}^F}, 1) \one_{\pi_{O_{is}^F} \in (0,1)}.$$
We assume that species occurrence indicators are conditionally independent Bernoulli random variables given prior occurrence probabilities, 
$$P(O_{is}^F = 1 \mid \pi_{O_{is}^F}) = \pi_{O_{is}^F}.$$ 
Posterior samples are provided by joint updating of $(O_{is}^F, \pi_{O_{is}^F})$ in a blocked Metropolis Hastings step. A similar procedure is used for plant occurrence indicators and occurrence probabilities, $(O_{is}^P, \pi_{O_{is}^P})$; full details are provided in the Supplement.

\subsubsection{Full model: covariate-informed latent factor network completion}

The likelihood in Equation \eqref{eq:Likelihood} describes the central probabilistic relationship governing the observation of the interaction. However, it cannot be fit directly due to its reliance on latent variables: the co-occurrence matrix, the true interaction matrix, and the detection probabilities. While we have already outlined the estimation of the co-occurrence matrix, it remains to describe the estimation of detection probabilities, and the central quantity of interest, true but unobserved interactions. To fit this model, we follow \cite{PapadogeorgouJASA} in estimating the latent variables via three submodels describing (1) interaction propensity, (2) detection probability, and (3) observable traits; the three submodels are linked by the likelihood and their shared dependence on unobservable latent species covariates. 

\begin{figure}[h!]
    \centering
    \includegraphics[width=\linewidth]{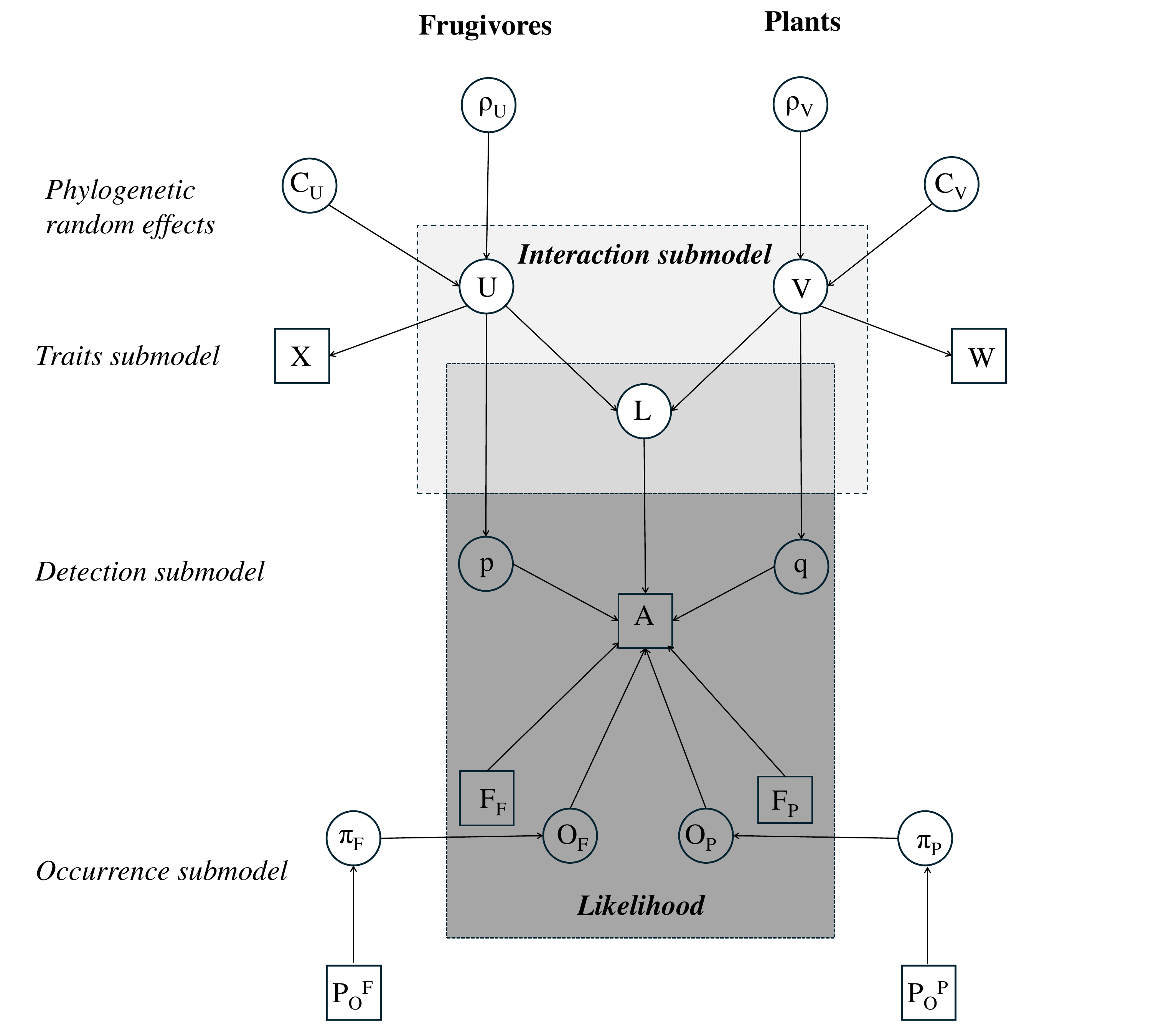}
    \caption{The Directed Acyclic Graph (DAG) of the core link prediction model. Observed data are represented as squares and estimated parameters as circles.}
    \label{fig:DAG}
\end{figure}

The complete link-prediction framework is illustrated in Figure \ref{fig:DAG}, with interaction, trait, detection, and occurrence submodels connected via latent species traits and observed interactions. A core \textit{interaction submodel} specifies that links $L$ between frugivores and plants are driven by unobservable latent traits $U,V$ of the respective species. A \textit{trait submodel} specifies the relationship between observable species covariates $X,W$ and latent traits, while a \textit{detection submodel} specifies the relationship between species detection probabilities $p,q$ and latent traits. Simultaneously, latent traits for each species are learned through a flexible phylogenetic random effects model, permitting learning of the influence of phylogenetic correlation $\bC_U, \bC_V$ and general interaction propensity from a limited and biased set of observed interactions and observable traits. The actual observability of recorded interactions $A$ is biased by study focus $\bF_F,\bF_P$ and the geographical location of studies, and hence the species occurrence $\bO_F, \bO_P$ in the study location, as captured by the \textit{likelihood}. An \textit{occurrence submodel} formalizes uncertainty in species occurrence and permits iterative sampling of species occurrence indicators given species occurrence probabilities $\pi_F, \pi_P$; these probabilities in turn depend on supplied prior probabilities $\bP_{O^F}$ and $\bP_{O^P}$. The process through which our model learns these latent covariates, and hence the true interaction networks, is described in detail below. 

To proceed with the development of the full modeling framework, we now assume that, in addition to the observed networks, study-level occurrence, and focus data, we have a (likely small) data set of known species traits. That is, for each frugivore species $i$ and each plant species $j$, we observe a vector of species-level traits, $\bX_i = (X_{i1}, X_{i2}, \cdots, X_{ip_M})^T$ and $\bW_j = (W_{j1}, W_{j2}, \cdots, W_{jp_P})^T$ respectively. The number of frugivore traits ($p_M$) and plant traits ($p_P$) may differ. Furthermore, we assume that we have phylogenetic correlation matrices for both frugivores and plants: $\bC_U, \bC_V$. 

Moreover, there exists a collection of unobserved and potentially unobservable traits for each species, which are nevertheless important for describing interaction patterns. 
For each frugivore species $i$ and each plant species $j$, we define an $H$-dimensional vector of latent species factors $\bU_i = (U_{i1}, U_{i2}, \cdots, U_{iH})^T$, and $\bV_j = (V_{j1}, V_{j2}, \cdots, V_{jH})^T$. These latent factors can be conceptualized as unobserved traits or preferences of the respective species, with $H$ providing an upper bound on the number of relevant traits. This $H$ is typically a small number as latent factors provide a low-dimensional summary of each species' ecological niche. Latent factors in this context are not identifiable due to rotational ambiguity and cannot be directly interpreted.

The phylogenetic correlation matrices, $\bC_U$ and $\bC_V$ provide the first source of information on latent factors. Phylogenetic dependence in the latent factors -- and hence in interactions -- is induced via phylogenetic structuring of a Gaussian process prior on latent factors. That is, for any index $h$ corresponding to a particular dimension of the latent space, the collection of all frugivore latent factors has a Gaussian process prior in which the covariance between the latent factors for any two frugivore species decreases with their distance on the phylogenetic tree, and similarly for plants: 
\begin{linenomath}
\begin{align} \label{eq:Factorprior}
\bU_{.h} \sim N(0, \Sigma_U),\quad
\bV_{.h} \sim N(0, \Sigma_V),
\end{align}
\end{linenomath}
where $\Sigma_U = \rho_U \bC_U + (1 - \rho_U)$, and similarly for $\Sigma_V$, and $\rho_U, \rho_V \in (0, 1)$ control the influence of phylogeny on latent trait structuring.

The observed trait matrices $\bX$ and $\bW$ provide the next source of information about latent traits. We assume that observed traits are driven by latent factors such that the observable covariates are each the manifestation of a particular combination of unobservable, ancestral traits. Concretely, for each frugivore trait $l$ and each plant trait $q$, and appropriate link functions $f_l$ and $g_q$, the \textit{trait submodels} represent observed traits via a generalized linear model on latent factors: 
\begin{linenomath}
\begin{align} \label{eq:Traits}
f_l^{-1}(E(\bX_{il} \mid \bU_i)) = \bbeta_{l0} + \bU_i'\bbeta_l,\quad
g_q^{-1}(E(\bW_{iq} \mid \bV_j)) = \bgamma_{q0} + \bV_j'\bgamma_q 
\end{align}
\end{linenomath}

More information about the latent factors is derived from the interaction patterns. Although latent factors are not identifiable, exploring what they \textit{might} represent aids us in developing an intuition for the model. In the context of a bipartite interaction network such as the Afrotropical Frugivory database, one could imagine that one latent frugivore trait could be a metabolic adaptation requiring a high-lipid content diet; a corresponding plant trait could be the lipid content of its fruit. A latent factor model learns such preferences and features in the absence of relevant trait data through observed interaction patterns, a process captured in the \textit{interaction submodel}: 
\begin{equation} \label{eq:Interaction}
    \text{logit}P(L_{ij} = 1 \mid \bX_i, \bU_i, \bW_j, \bV_j) = \lambda_0 + \sum_{h=1}^H \lambda_h U_{ih}V_{jh}
\end{equation}

The interaction submodel specifies the propensity for interaction between frugivore species $i$ and plant species $j$ as a function of a baseline interaction score $\lambda_0$ and the weighted inner product of their respective latent factors. Interaction propensities are independent of observed traits conditional on the latent factors: traits impact interactions only through the latent factors. The weights $\lambda_1, \cdots, \lambda_H$ are modeled by a multiplicative inverse gamma shrinkage process of \cite{Bhattarcharya}, so that all weights are positive and the weights stochastically decrease with the index $h$. The motivation for this construct is to induce shrinkage, decreasing the contribution of higher index latent factors to the interaction propensity so that only a small number of latent factors prove important.

The final submodel specifies the relationship between species detection and latent factors. We define the detection probability for frugivore species $i$, $p_i$ to be the probability of detecting an interaction involving frugivore $i$ and a perfectly detectable plant partner, given that it occurs in the study period. Similarly, the probability of plant detection $q_j$ gives the probability of detecting an interaction involving a plant $j$ and a perfectly detectable frugivore. As specified in the likelihood in equation \eqref{eq:Likelihood}, the interaction probabilities for frugivores and plants are independent, such that the probability of observing an interaction involving a frugivore $i$ and a plant $j$ is simply the product $p_i q_j$. The \textit{detection submodels} express that latent factors drive the detection probabilities for each frugivore $i$ and each plant $j$: 
\begin{linenomath}
\begin{align} \label{eq:Detection}
E[\text{logit}(p_i) \mid \bU_i, \bX_i]  = \delta_0 + \bU_i'\bdelta,\qquad
E[\text{logit}(q_j) \mid \bV_j, \bW_j]  = \zeta_0 + \bV_j'\bzeta 
\end{align}
\end{linenomath}

The full model above is given a Bayesian specification. Since the posterior distribution is complex, we follow standard practice in relying on Markov chain Monte Carlo sampling algorithms for model fitting. Due to the logistic regression form of the conditional likelihoods, a Gibbs sampler can be used after Polya-Gamma data augmentation \citep{Polson01122013}. Details on the full Bayesian model specification including all priors are provided in the Supplementary Materials.

We fit our link prediction model to the Afrotropical Frugivory database and supporting data described in Sections \ref{sec:AF_data} and \ref{sec:AF_supplemental data}. We consider both the basic \textbf{Co}variate-\textbf{I}nformed \textbf{L}ink Prediction model (COIL), which does not sample occurrence probabilities, and the expanded model including sampling of occurrence probabilities (COIL+). We fit these models under the following scenarios: (i) a naive interaction-based occurrence construction that assigns to all species not directly observed in the study a 0\% prior probability of occurrence (COIL 0/100, COIL+ 0/100) and (ii) an expert-defined prior occurrence probability structure using the method described in Section \ref{ssec:model}, where the prior probability of occurrence is determined based on the proximity of the study to other studies in which the species was observed (COIL Exp, COIL+ Exp). The same construction is used for both frugivores and plants in all scenarios. Further details on prior specification under each scenario are provided in the Supplementary Materials, including a discussion of a recommended default prior occurrence construction. Posterior samples are obtained across four independent MCMC chains, each run with 20,000 iterations, including a burn-in of 10,000 iterations and 10,000 post-burn-in iterations thinned to 5\% to reduce the memory footprint of the high dimensional posterior probability array while still providing sufficient chain length to ensure convergence.

\section{Results}

In this section, we present a summary of model performance under the various scenarios developed for the Afrotropical Frugivory database. MCMC diagnostics for scalar parameters show good mixing, suggesting convergence to the stationary distribution; see Supplementary Materials for details. 

\begin{figure}[htp]
    \centering
    \includegraphics[width=0.72\linewidth]{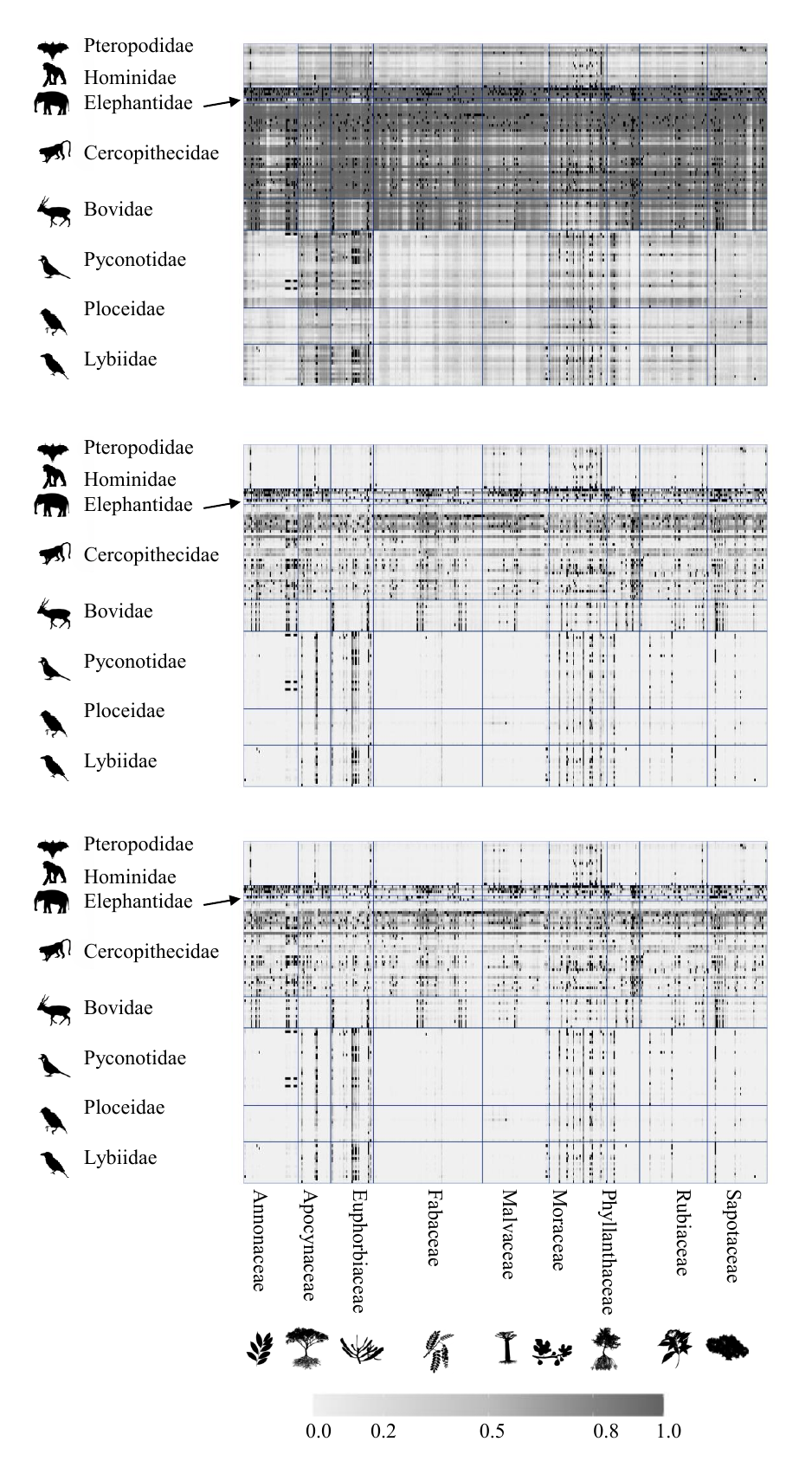}
    \caption{Heatmaps showing seed dispersal interactions in the posterior network imputed using COIL 0/100  (top), COIL Exp (middle) , and COIL+ Exp (bottom). Black rectangles indicate recorded interactions. Silhouettes from \textit{PhyloPic} \citep{PhyloPic2025}. }
    \label{fig:heatmaps_main}
\end{figure}

The imputation of latent interaction probabilities significantly increases the estimated prevalence of frugivory interaction, or the proportion of all possible frugivore-plant pairs that would interact given the opportunity, from 2.3\% based on direct observation to 4.5\%-4.6\% based on imputation under the expert-defined occurrence prior, as can be seen in the more dense imputed heat maps presented in Figure \ref{fig:heatmaps_main}. In addition to the 5,936 known frugivory interactions recorded in Afrotropical forests, we predicted 5,637 - 5,897 likely unrecorded interactions, where ``likely interactions'' were those with a posterior mean interaction probability greater than 75\%. Both model specifications under the expert-defined occurrence prior yield superior performance relative to the interaction-based occurrence construction, with substantial gains for the extended model. 

\begin{figure}[!htp]
    \centering
    \includegraphics[width=0.5\textwidth]{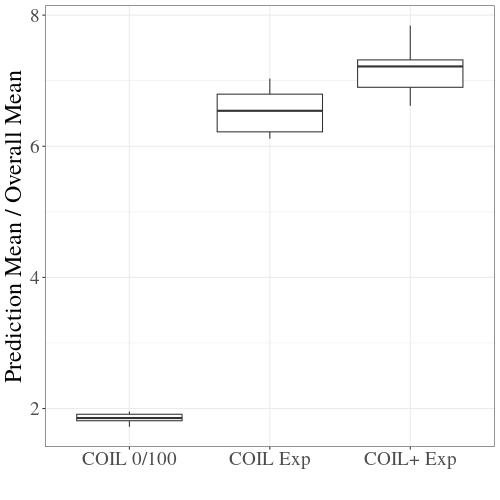}
    \label{fig:oos_precsision}
    \caption{Out of sample performance under the three modeling scenarios as measured by pseudo-precision or the ratio of posterior mean interaction probability in heldout true interactions to the posterior mean overall.}
	\label{fig:oos_recallandprecision}
\end{figure}

Validation of missing link imputation for presence-only data presents a challenge because true negatives are unknown: non-edges may occur due to the absence of a true interaction, non-overlap of species or relevant phenologies, or incomplete observation. One traditional metric available in this context is simply the recall: we find that COIL 0/100, which does not allow uncertainty in occurrences, performs the best from this perspective, recovering 81\% of held out true interactions with a posterior probability greater than 75\%. In the scenario of expert-defined events, the recall is 57\% for COIL+ Exp and 58\% for COIL Exp. The obvious limitation of this metric is that it does not consider the plausibility of predicted prevalence in each model.

Indeed, it is of key importance to verify that a model is able to recover true edges without simply predicting interactions for all pairs. In this context a useful cross-validation scheme witholds only known interactions and computes \textit{pseudo-precision} as the ratio of the mean posterior probability among the known interactions held out to the mean posterior probability overall \citep{PapadogeorgouJASA}. Defined in this way, pseudo-precision should be in the interval $(1, 1/\text{True Prevalence})$, with larger values generally indicating a superior discrimination of true edges. Under the expert-defined occurrence scenario, we obtain a pseudo-precision of 7.17 for COIL+ Exp  and 6.55 for COIL Exp, indicating strong ability to identify true interactions, with nontrivial gains for the extended model. Using interaction-based occurrence data only, in contrast, we obtain a markedly lower pseudo-precision of 1.85 for COIL 0/100, suggesting this model specification results in over-prediction of interactions. 

\section{Discussion}
Examining the posterior network heatmaps (Figure \ref{fig:heatmaps_main}), we see that all models increase the prevalence of interactions relative to the observed network, revealing many potential links not yet documented in the literature. In the COIL+ model, the 5,637 predicted links correspond to a median of nine extra interactions per frugivore across 957 plant species—a modest but biologically plausible increase. These new links are concentrated among poorly studied species, thereby reducing bias from unequal sampling intensity. This is reflected in the decline in correlation between sampling effort and estimated interaction richness, from 0.88 in the observed network to 0.43 in the posterior network (thresholded at 75\% interaction probability to represent very likely interactions).

Comparing the imputed networks, there is a much greater density of high posterior probability interactions in the network imputed without sampling occurrence indicators or occurrence probabilities (COIL 0/100) compared to the models which allow uncertain occurrences with expert-defined probabilities (COIL Exp, COIL+ Exp). Defining likely interactions to be those with posterior probability greater than 75\%, we find an implausibly high posterior interaction prevalence of 28.2\% using COIL 0/100 versus a posterior interaction prevalence of 4.4\% and 4.6\% under COIL Exp and COIL+ Exp respectively. Hence, both COIL Exp (sampling occurrence indicators) and COIL+ Exp (sampling occurrence indicators and probabilities) suggest a more moderate interaction prevalence, with COIL Exp yielding slightly greater interaction density. Species vary in their degree of specialization and, therefore, niche differentiation (e.g., \cite{galetti_trophic_2016}). However, in general, wild animal diets tend to be specialized \citep{hutchinson_dietary_2022}, making extremely high prevalence of interactions unlikely. Poor pseudo-precision for COIL 0/100 is consistent with the above finding that, while the model has an apparent advantage in terms of recall, this is the result of over-estimating interactions across the board, as evidenced by the model's poor pseudo-precision.

To further assess the differences in model performance, we can examine the posterior mean occurrence probabilities estimated for species not directly observed. Table \ref{tab:occ-table} summarizes the estimated occurrences for plants and frugivores in each modeling scenario. Because COIL 0/100 does not permit updating of occurrences, treating all unobserved species as absent, all edges involving unobserved species are non-informative; this results in the model learning excessively high interaction propensities across the board. Under COIL Exp and COIL+ Exp, we see more plausible posterior-occurrence probabilities, strongly influenced by the expert-defined prior. Hence, both of these models treat non-edges involving species that were not directly observed in a given study as more or less informative depending on expert defined occurrence probabilities. These in turn depend on the presence of that species in studies at different levels of proximity to the study in question, for example, estimating that a plant species observed previously at the same site has a 68\% to 69\% posterior mean probability of occurrence under COIL Exp, COIL+ Exp respectively.

A key difference between these two models is that COIL Exp does not update occurrence probabilities, so it does not sample occurrence indicators for any species given a 0\% occurrence probability at a given site. COIL+ Exp, in contrast, does, producing a substantially higher mean posterior occurrence probability of 8\% for both frugivores and plants in that scenario. This difference explains the slightly higher density of interactions in the posterior network for COIL Exp compared to COIL+ Exp, and the slightly worse performance of COIL Exp with respect to pseudo-precision. 

\begin{table}[htp]
\begin{tabular}{@{}lrrrrrr@{}}
\toprule
 & \multicolumn{3}{c}{\textbf{Plant Taxa}}           & \multicolumn{3}{c}{\textbf{Frugivore Taxa}} \\ \midrule
                & COIL 0/100 & COIL Exp & COIL+ Exp                 & COIL 0/100     & COIL Exp    & COIL+ Exp    \\ \midrule
Different region   & 0.00       & 0.00     & \multicolumn{1}{r}{0.08} & 0.00           & 0.00        & 0.08        \\
Same region only    & 0.00       & 0.23     & \multicolumn{1}{r}{0.24} & 0.00           & 0.25        & 0.24         \\
Same country only & 0.00       & 0.46     & \multicolumn{1}{r}{0.47} & 0.00           & 0.50       & 0.48         \\
Same site         & 0.00       & 0.68     & \multicolumn{1}{r}{0.69} & 0.00           & 0.75        & 0.72         \\
Same study        & 1.00       & 1.00     & \multicolumn{1}{r}{1.00} & 1.00           & 1.00        & 1.00         \\ \bottomrule
\end{tabular}
\caption{Posterior mean occurrence probabilities for plant and frugivore taxa across the three models, assuming the taxa has been observed in an interaction no closer than (from top to bottom) : a different region in Africa, the same region in Africa, the same country, the same site, and the same study.}
\label{tab:occ-table}
\end{table}

The heatmaps (Figure \ref{fig:heatmaps_main}) show strong phylogenetic influence, visible as horizontal and vertical strips of high posterior probability interactions for similar frugivore and plant species. In fact, we obtain high posterior mean values for the phylogenetic structuring coefficients, with 95\% credible intervals of $(0.91, 0.95)$ and $(0.91, 0.94)$ for $\rho_U$ and $\rho_V$, which describe the phylogenetic influence on frugivore and plant latent factors, respectively. Frugivores and plants share a deep evolutionary history, and fleshy fruits evolved across many angiosperm families \citep{eriksson_evolution_2016}. Our results highlight the importance of phylogeny for frugivory, coinciding with other research (e.g., globally, phylogeny predicts vertical foraging in mammals and birds \citep{jantz_functional_2024}).

A very useful feature of the proposed framework is that it permits trait matching, or the identification of physical trait compatibilities influencing frugivory. For frugivores, we consider generation length and log body mass and, for plants, we consider fruit length, fruit width, and wood density, as described in Section \ref{sec:AF_supplemental data}. All trait matching analyses are performed using the posterior network obtained using the COIL+ Exp specifications. 

\begin{figure}[htp]
    \centering
    \includegraphics[width=0.75\linewidth]{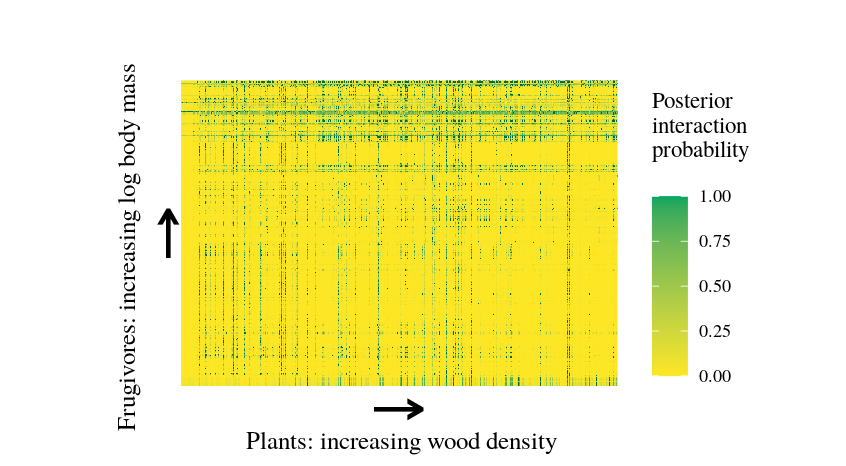}
    \caption{COIL+ Exp posterior network organized by the  most important traits for frugivores and plants. Interaction propensity generally increases from bottom to top with log body mass, and decreases from left to right with wood density. }
    \label{fig:heatmap_with_traits}
\end{figure}

First, we evaluate variable importance using the resampling method proposed by \cite{PapadogeorgouJASA}: for each frugivore trait $X_{.l}$, we compute the squared correlation between the trait (across all frugivore species) and the logit interaction probability with each plant species, and then average over posterior samples and plant species to obtain an observed correlation test statistic. Next, we permute the trait values and repeat the process for a large number ($B=100$) of bootstrap iterations. Finally, we define variable importance to be the number of standard deviations by which the observed test statistic differs from the permuted test statistic. We perform a similar permutation test for plant traits $W_{.q}$. Using this procedure, we find increasing log body mass and wood density to be the most important variables for frugivores and plants, respectively. In Figure \ref{fig:heatmap_with_traits}, we see the greatest density of high posterior probability interactions appears for medium wood density plants and high body mass animals. Another notable cluster of high posterior probability interactions occurs for low to moderate body mass animals with low wood density plants.

Although our analysis covers a large geographic area, mutualistic interactions are moderated by local climate and biogeography \citep{McFadden}. As such, we should not necessarily expect a strong global pattern to emerge in trait matching, but instead expect to see strong heterogeneity in association between traits and interaction propensity. Because it is likely that the frugivore traits which facilitate frugivory with a given plant will vary by the plant of interest (and similarly for plant traits), we consider an additional trait matching procedure which permits analysis of this hypothesized heterogeneity. 
Whereas the variable importance measure above considers the mean squared correlation across all pairs, the procedure below computes a signed correlation for each species. For the $l^{th}$ frugivore trait, we proceed by taking the vector of observations of this trait for all frugivore species $X_{.l}$. Next, for each plant species and each posterior sample $r$, we compute the correlation between the trait vector $X_{.l}$ and the vector of logit interaction probabilities with respect to the $j^{th}$ plant species, $P_{.jr}$. We repeat the process for each posterior sample to obtain a posterior mean correlation coefficient for each trait with each plant species; this allows us to see which frugivore traits are most positively and negatively associated with interactions with each plant species. We perform a similar procedure for plant traits.

\begin{figure}[htp]
    \centering
    \includegraphics[width=0.85\linewidth]{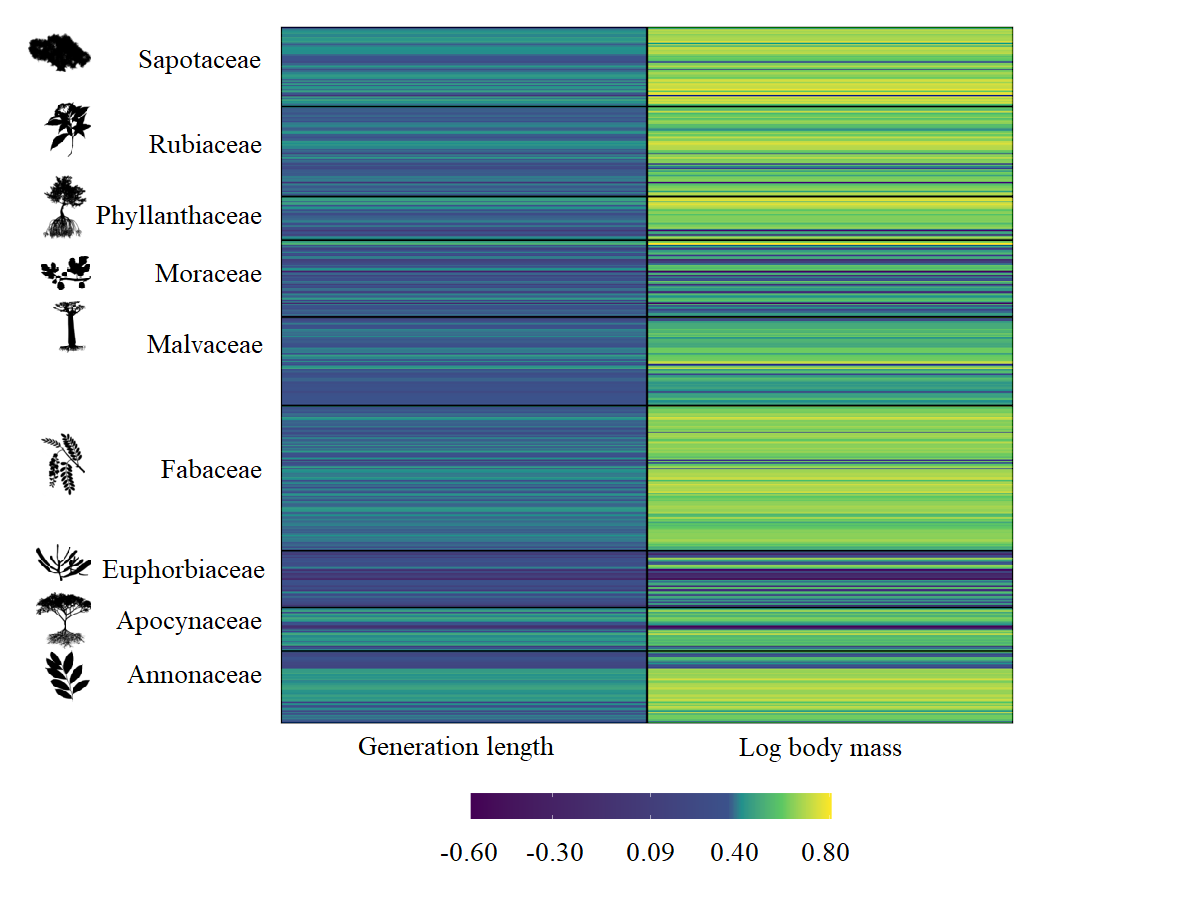}
    \caption{Correlation of continuous frugivore covariates with posterior interaction propensity with the respective plant species, grouped by family.}
    \label{fig:enter-label}
\end{figure}

 The frugivore trait log body mass is found to be the most important predictor of frugivory across plant species. Although gape size (rather than body size) is often used for trait matching \citep{PapadogeorgouJASA}, data on gape size are less widely available. In such cases, log body mass—which is expected to correlate with gape size \citep{moran_can_2010}—can serve as a useful proxy, as it does here. Looking at the correlation between log body mass and interaction propensity by plant family, we see that it is generally positive, consistent with the findings of a positive relationship between body mass and fruit removal rates in birds \citep{Munoz2017}. Notable exceptions are concentrated in the families Euphorbiaceae (spurges) and Moraceae (figs), for which there is a negative correlation between animal log body mass and frugivory; interestingly, both families are characterized by a milky latex. In general, we see weaker and more negative correlations between generation length of the animal and frugivory across plant species, with larger positive correlations for the plant families Annonaceae and Sapotaceae. One potential explanation for this trend is that species with shorter generation lengths, which invest more in reproduction than survival, can exploit a greater variety of fruit resources because they may be better adapted to variable environments.

\begin{figure}[htp]
    \centering
    \includegraphics[width=0.9\linewidth]{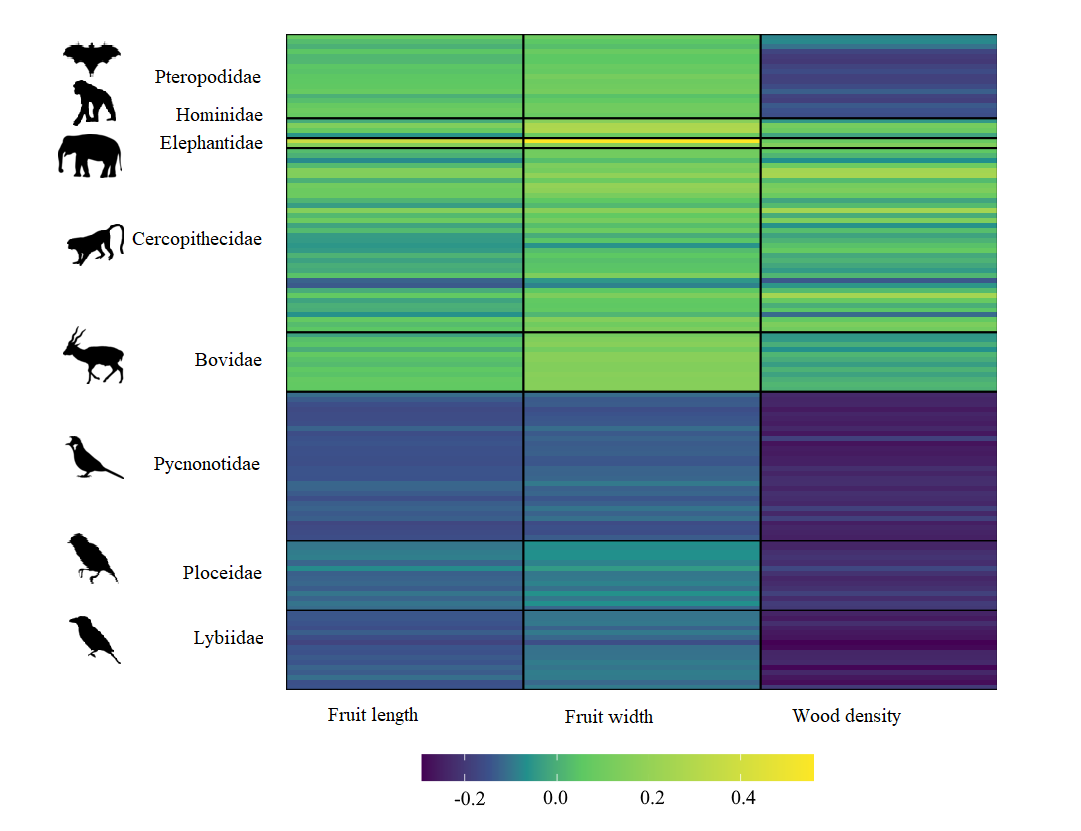}
    \caption{Correlation of plant covariates with posterior interaction propensity with the respective frugivore species, grouped by family.}
    \label{fig:enter-label}
\end{figure}

Across all frugivore species, the plant trait most strongly associated with interaction propensity (as measured by the squared correlation) is wood density. For most frugivore species, this correlation is negative, with notable exceptions in the families Cercopithecidae (old world monkeys), Hominidae (apes), and Elephantidae (elephants) and, to a lesser degree, Bovidae (ruminants). It is not surprising that, aside from these exceptions, the relationship between wood density and frugivory tends to be negative: high wood density trees invest more in their woody structure and grow more slowly, perhaps with fruit that is nutrient dense but having lower production. In contrast, low wood density species grow more quickly, producing more fruit but potentially fruit that is less nutritionally dense. Additionally, species with high wood density may be more dependent on the right environmental conditions to produce fruit, providing a reliable food source that may be more readily available to many animals, whereas low wood density trees may be less dependent on particular environmental conditions for producing fruit \citep{lima_phenology_2010}. Alternately, it may simply be that wood density is in part a proxy for fruit which is prohibitively large to the flying species for which the correlation is negative, yet easily accessible to larger or more dexterous mammals.

Considering fruit size directly, we find fruit length of comparable importance to fruit width, but with larger positive correlations with fruit width. This contradicts previous research suggesting that fruit width is more predictive of frugivory than fruit length \citep{yu_contrasting_2024}.  For both fruit length and width, we see negative correlations for animals in the avian families Pycononitidae (bulbuls), Ploceidae (weavers), and Lybiidae (barbets) and weaker positive correlations for animals in the families Bovidae and Pteropodidae (bats). The pattern is similar for fruit width, but with larger positive correlations, notably for the families Bovidae and Cercopithecidae. There is ample precedence in the literature that bird families tend to consume smaller fruits while other families consume larger fruits; for example, \cite{Florchinger2010} shows that, in Kenya, birds fed on smaller fruits whereas primates fed on larger fruits. Notably, bats show a similar pattern to birds in terms of interaction propensity being negatively related to wood density, but an opposite pattern with respect to fruit size, with weaker positive correlations between fruit length and width and interaction propensity for bats. These results support previous research; for example, considering a Panamanian fig community, birds tend to disperse smaller fruits whereas bats disperse fruits from a wide range of size classes \citep{kalko_relation_1996}. Size matching between birds/ bats and the fruits they consume is likely driven by fruit handling ability \citep{rojas_combination_2021}.

\subsection{Conclusions}

We introduced a framework that incorporates uncertainty in species occurrence into latent factor–based link prediction while accounting for phylogeny and observable traits. Our method improves on existing approaches by (1) generating informative priors for occurrence probabilities through borrowing information across studies, and (2) updating those probabilities via an additional sampling step. These innovations reduce bias when interaction data are drawn mainly from taxonomically biased sources, and prevent the implausibly dense posterior networks that can arise if occurrence uncertainty is ignored.

Applied to a new database of Afrotropical frugivory interactions spanning 957 plants and 267 vertebrates, our approach suggested over 5,000 previously undocumented links, concentrated among poorly studied species. By reducing the correlation between sampling effort and estimated interaction richness, the method mitigates a key source of bias and enables more balanced inference across taxa. We also introduced a trait-matching approach based on the posterior networks, which accommodates heterogeneity in trait–frugivory associations and reveals both novel and established patterns in the Afrotropical frugivory data.

This framework opens the door to richer models of ecological networks, while the phylogenetically structured latent space approach facilitates out-of-sample prediction, including for species entirely lacking interaction data. Future extensions include incorporating spatiotemporal variability in interaction propensity and integrating phenological information on frugivore presence and plant fruiting. Ecological research increasingly recognizes that interaction networks are not static but vary through space and time due to local trait distributions, abundances, environmental conditions, and even the influence of non-interacting species \citep{Poisot, Gravel2019}. In the absence of suitable covariates, spatiotemporal latent factor models can efficiently capture such variability \citep{durante_duson2014}. Incorporating phenological models where available would further expand this framework. Together, these extensions would enable more accurate and dynamic representations of ecological interactions.

\subsection*{Data availability}
 Data and code to reproduce the tables and figures are also available at \url{https://github.com/jennifernoelle/COILplus}. For the Afrotropical Frugivory Database, species-level taxonomic information has been replaced with generic taxa labels in accordance with data-sharing restrictions.

\subsection*{Acknowledgments}
This project received funding from the European Research Council (ERC) under the European Union’s Horizon 2020 research and innovation programme (grant agreement No. 856506; ERC Synergy project LIFEPLAN) and from the U.S. National Science Foundation under Award No. 2426762 (Project Title: NSF-AoF: III: Small: Autonomous biodiversity monitoring through wireless communication technologies and artificial intelligence). We thank Clémentine Durand-Bessart for sharing frugivory interaction data and Georgia Papadogeorgou and Charles Nunn for valuable discussions and comments on earlier versions of this manuscript.

\subsection*{Conflict of interest}
The authors declare no conflict of interest.

\subsection*{Authors' contribution}
Jennifer Kampe, Camille DeSisto, and David Dunson jointly designed the method. Jennifer Kampe led the writing of the manuscript, implemented the model, developed the software, and analyzed the data. Camille DeSisto assembled the dataset, which includes both original data collected as part of her other research and additional data compiled from a literature review. All authors contributed critically to the drafts and approved the final version for publication.

\clearpage
\bibliographystyle{plainnat}
\bibliography{references}

\clearpage
\appendix
\section*{Supplementary Material}

\section{Posterior computation}

Given the sensitivity of the COIL model to uncertainty in occurrence under extreme taxonomic bias, the proposed method introduces the following extension (COIL+): (i) the user submits occurrence probabilities, which are assumed to be known with uncertainty; (ii) these probabilities are used as the center parameter in a prior for the actual occurrence probabilities, which are sampled at each iteration; and (iii) the sampled values of the occurrence probabilities are used in imputing the occurrence indicators. We present the method in detail below using the frugivore occurrences $O_{is}^F$ and occurrence probabilities $\pi_{O_{is}^F}$,  for illustration, however the same procedure is followed for plant $O_{is}^P$ occurrences and occurrence probabilities $\pi_{O_{is}^P}$. 

Under the COIL framework \citep{PapadogeorgouJASA}, the user provides a list of probabilities of occurrence which are assumed to be \textit{known}. When a species $i$ is observed interacting with any other species in study $s$, its occurrence is known to be one, $O_{is}^F = 1$. However, if a species is not observed in any interaction in the study $s$, $\sum_{j} A_{ijs} = 0$, its occurrence is unknown and it is sampled as follows. 

The prior occurrence probability is: 
$$P(O_{is}^F = 1 \mid \pi_{O_{is}^F}) = \pi_{O_{is}^F},$$
and the likelihood contribution is 
$$P(A_{ijs} \mid .) \propto \prod_{ijs: F_{ijs}O_{ijs}L_{ij} = 1} (p_i q_j)^{a_{ijs}}(1- p_i q_j)^{1-a_{ijs}}  \prod_{ijs: F_{ijs}O_{ijs}L_{ij} = 0} \one(a_{ijs} = 0).$$
Posterior updates are obtained via a Gibbs sampling step with the following full conditional: 
\begin{equation}
\begin{aligned}\label{eq:SM_likelihood}
P(O_{is}^F = 1 \mid . ) &\propto \pi_{O_{is}^F} \prod_{j: F_{ijs}O_{js}L_{ij} = 1} (p_i q_j)^{a_{ijs}}(1- p_i q_j)^{1-a_{ijs}}   \prod_{j: F_{ijs}O_{ijs}L_{ij} = 0} \one(a_{ijs} = 0) \\
 &= \pi_{O_{is}^F} \prod_{j: F_{ijs}O_{js}L_{ij} = 1} (1- p_i q_j)^{1-a_{ijs}}  \\
 &=  \pi_{O_{is}^F} \prod_{j} (1- p_i q_j)^{F_{ijs}O_{js}L_{ij}}.
\end{aligned}
\end{equation}

In the COIL+ approach, we use the user-provided probabilities as the center for the prior on occurrence probabilities, which are updated
in the sampler. These values either correspond to an educated guess about the occurrence probabilities based on available information or one can use the following default values: 
$$p_{O_{is}^F} = \begin{cases} 1 & \text{ if } \sum_{j} a_{ijs} > 0 \\
0.75 & \text{otherwise} \end{cases}.$$
When $\sum_{j} a_{ijs} > 0$, some interaction involving species $i$ is observed in study $s$, so species $i$ is known to have occurred in study $s$; neither the occurrence probability nor the occurrence indicator is sampled in this case. Otherwise, the above probabilities are used as the prior modes for occurrence probabilities, given a weakly informative truncated normal prior:

\begin{equation}\label{eq:SM_piprior}
    \pi_{O_{is}^F} \sim f_O = \Nf(p_{O_{is}^F}, 1) \one_{\pi_{O_{is}^F} \in (0,1)},
\end{equation}
Posterior samples are obtained by block updating of $O_{is}^F, \pi_{O_{is}^F}$. In particular, at each iteration we independently sample new values of $O_{is}^F, \pi_{O_{is}^F}$ using the following proposal distributions:
\begin{align}
   {O_{is}^F }^* & \leftarrow \text{Bernoulli}(p^*)  \label{eq:sm_proposal_pi} \\ 
  \pi_{O_{is}^F }^* & \leftarrow \Nf(\pi_{O_{is}^F}^{(t-1)}, 0.1), \label{eq:sm_proposal_O}  
\end{align}

with $$p^* = \begin{cases} p_{0,1} \text{ if } {O_{is}^F }^{(t-1)} = 0 \\
1-p_{1,0} \text{ if } {O_{is}^F }^{(t-1)} = 1, 
\end{cases}$$
where $p_{0,1}$ and $p_{1,0}$ are the probability of switching the indicator from a 0 to a 1 and from a 1 to a 0 respectively. Then we either accept or reject the proposed pair of new values jointly in a Metropolis Hastings step.

\begin{algorithm}[H]
\caption{Blocked Metropolis Hastings: occurrence indicators and probabilities}\label{alg:cap}
\begin{algorithmic}
\Inputs{${O_{is}^F}, p_{O_{is}^F}$}
\Initialize{$O_{is}^F \gets \text{Bernoulli}( p_{O_{is}^F}), \pi_{O_{is}^F}^{(0)}  \gets \Nf( p_{O_{is}^F}, 1)$}
\For{t = 1 to T}
    \State Update all other parameters from their full conditionals
    \State $\pi_{O_{is}^F}^* \gets \Nf(\pi_{O_{is}^F}^{(t-1)}, 0.1)$
    \State ${O_{is}^F}^* \gets \text{Bernoulli}(p^*)$
    \State $\alpha \gets \frac{f_O({O_{is}^B}^* \mid \pi_{O_{is}^F}^*, \cdots )f_\pi(\pi_{O_{is}^F}^*) g_O({O_{is}^F}^{(t-1)}| {O_{is}^F}^*)g_\pi(\pi_{O_{is}^F}^{(t-1)}| \pi_{O_{is}^F}^*)}{f_O({O_{is}^B}^{(t-1)} \mid \pi_{O_{is}^F}^{(t-1)}, \cdots )f_\pi(\pi_{O_{is}^F}^{(t-1)}) g_O({O_{is}^F}^{*}| {O_{is}^F}^{(t-1)})g_\pi(\pi_{O_{is}^F}^{*}| \pi_{O_{is}^F}^{(t-1)})}$
    \State $u \gets \text{Uniform}[0,1]$
        \If{$\alpha \leq u$}
            \State ${O_{is}^F}^{(t)} \gets {O_{is}^F}^*$
            \State $\pi_{O_{is}^F}^{(t)} \gets \pi_{O_{is}^F}^*$
        \Else 
            \State ${O_{is}^F}^{(t)} \gets {O_{is}^F}^{(t-1)}$
            \State $\pi_{O_{is}^F}^{(t)} \gets \pi_{O_{is}^F}^{(t-1)}$
        \EndIf
\EndFor
\end{algorithmic}
\end{algorithm}

Here, $f_\pi$ is given by Equation \eqref{eq:SM_likelihood}, $f_O$ is given by Equation \eqref{eq:SM_piprior}, and $g_O, g_\pi$ are the asymmetric proposals described in Equations\eqref{eq:sm_proposal_pi} and \eqref{eq:sm_proposal_O}.
We recommend switching probabilities $p_{0,1} = 0.25$ and $p_{1,0} = 0.65$ to slightly favor non-occurrence, consistent with geographically limited ranges in the present application. These values and the Metropolis step size of $0.1$ were selected to yield an acceptance ratio of approximately 1/3. This simple blocked sampler yielded superior mixing and recovery of simulated data relative to other possible sampling schemes considered, including sequential sampling of occurrence indicators and occurrence probabilities, blocked sampling using data augmentation, and adaptive Metropolis. The remaining sampling steps are as used in \cite{PapadogeorgouJASA}.

\section{Application Details}

\subsection{Data Preparation}

The physical characteristics of the fugivore and the plant are collected by performing a literature search as detailed in \cite{DeSistoKampe}. The resulting data are nearly complete for frugivores with only two species missing generation lengths; we impute these missing values using genus and family means. Plant data are less complete, with 107 observations missing fruit width, 94 missing fruit length, and two missing wood density. Where only fruit length is available, we impute $width = length/ mean(length:width)$. If fruit length is not available, we use genus or family means (with genus preferred if available). Genus and family means are used for missing values of wood density.

\subsection{Prior Elicitation}

In the African frugivory case study, we consider three priors for the occurrence probabilities $\pi_{O_{is}^F}$ and  $\pi_{O_{is}^P}$. First, we consider a naive definition of occurrence based on interaction, using a prior in which species not directly observed in an interaction in the study are given a prior probability of occurrence 0\%, and those observed are given a prior probability of occurrence 100\%. Next, we propose a default prior to be used in the absence of expert information, in which species not directly observed in the study are given a 75\% prior probability of occurrence (\textit{75/100}); this construction yielded the best performance on a grid search of similar \textit{XX/100} priors. Finally, we consider an expert-defined prior in which the prior probability of occurrence is determined based on the study's proximity to other studies in which the species was observed (\textit{Expert}). The same construction is used for both frugivores and plants in all scenarios, summarized in Table \ref{tab:SM_priors}. 

\begin{table}[htp]
\centering
\begin{tabular}{@{}rlrrr@{}}
\toprule
  &                   & 0/100 & 75/100 & Expert  \\ \midrule
1 & Different zone    & 0.00   & 0.75      & 0.00  \\
2 & Same zone only    & 0.00   & 0.75      & 0.25 \\
3 & Same country only & 0.00   & 0.75      & 0.50 \\
4 & Same site         & 0.00   & 0.75      & 0.75 \\
5 & Same study        & 1.00   & 1.00      & 1.00  \\ \bottomrule
\end{tabular}
\caption{Prior occurrence probabilities under the three scenarios. Zone refers the geographical region of Africa in which the study was conducted: Central, Eastern, or Western. }
\label{tab:SM_priors}
\end{table}

We find these priors to be strongly influential on data analysis results and hence suggest careful selection of prior occurrence probabilities. Running the COIL and COIL+ model under the scenarios described above, we obtain the following posterior probabilities of occurrence for frugivores and plants, in Figures \ref{tab:SM_postocc_frugs} and \ref{tab:SM_postoccs_plants}, respectively. Posterior occurrence probabilities under the expert-defined scenario are generally close to the prior mode, both in COIL and in COIL+, with the exception of the category corresponding to species never observed in a given zone of Africa. In this case, we see substantial movement from the prior mode under the COIL+ 
model, from a 0\% prior probability to a 10\% posterior probability of occurrence. The default 75/100\% prior yields ecologically implausible occurrences with a mean posterior occurrence probability of approximately 0.71-0.75 (regardless of sampler) for plants that have never even been observed in the same zone of Africa, and similarly for frugivores. Nevertheless, model performance does not suffer excessively under this proposed default prior, as we detail below.

\begin{table}[htp]
\resizebox{\columnwidth}{!}{%
\begin{tabular}{@{}lrrrrr@{}}
\toprule
                    & COIL 0/100 & COIL 75/100 & COIL+ 75/100 & COIL Exp & COIL+ Exp \\ \midrule
Different zone & 0.00 & 0.75 & 0.74 & 0.00 & 0.08 \\ 
 Same zone only & 0.00 & 0.75 & 0.74 & 0.25 & 0.24 \\ 
Same country only & 0.00 & 0.75 & 0.73 & 0.50 & 0.48  \\ 
Same site & 0.00 & 0.75 & 0.73 & 0.75 & 0.72 \\ 
Same study & 1.00 & 1.00 & 1.00 & 1.00 & 1.00  \\                  \bottomrule
\end{tabular}
}
\caption{Posterior mean occurrence probabilities for frugivores under the various scenarios.}

\label{tab:SM_postocc_frugs}
\end{table}

\begin{table}[htp]
\resizebox{\columnwidth}{!}{%
\begin{tabular}{@{}lrrrrr@{}}
\toprule
    & COIL 0/100 & COIL 75/100 & COIL+ 75/100 & COIL Exp & COIL+ Exp  \\ \midrule
Different zone & 0.00 & 0.74 & 0.74 & 0.00 & 0.08 \\  Same zone only & 0.00 & 0.74 & 0.74 & 0.23 & 0.24  \\ 
Same country only & 0.00 & 0.73 & 0.73 & 0.46 & 0.47 \\ 
Same site & 0.00 & 0.72 & 0.71 & 0.68 & 0.69  \\ 
Same study & 1.00 & 1.00 & 1.00 & 1.00 & 1.00 \\ \bottomrule
\end{tabular}
}
\caption{Posterior mean occurrence probabilities for plants under the various scenarios.}

\label{tab:SM_postoccs_plants}
\end{table}

\subsection{Model performance}

Due to the presence-only nature of nearly all interaction data, including that in the African Frugivory dataset, it is not straightforward to provide a quantification of the recall/pseudo-precision trade-off posed by the competing models. As such, we recommend model selection be guided by a thoughtful evaluation of the plausibility of prior and posterior occurrence probabilities, pseudo-precision, recall, and the loss function of the practitioner.

The analysis of pseudo-precision should be done in the context of expectations about interaction prevalence. We define pseudo-precision as the ratio of the mean posterior probability of interaction for held-out known edges to the ratio of the mean posterior probability overall, and so, assuming that not all edges are truly possible, pseudo-precision greater than one indicates good discrimination of edges from non-edges.  However, higher values of pseudo-precision are not always desirable. Pseudo-precision of a perfect model should be approximately equal to the reciprocal of interaction prevalence overall; this provides a reasonable upper bound for pseudo-precision. Hence, evaluation of pseudo-precision should be guided by expert knowledge on interaction prevalence, with desirable values falling in the interval $(1, 1/\text{True Prevalence}).$ For example, while specialization varies between species, we might assume that many frugivores are opportunistic, and hence any given frugivore would likely eat at least of 10\% of fruit species if given the chance. In that case, this lower bound on prevalence indicates that $100/10 = 10$ would be a rough upper bound for pseudo-precision. 

We define recall to be the proportion of true heldout interactions having posterior probability greater than a given threshold (50\% and 75\% in this application). Depending on the intended use for the posterior link prediction probabilities, users can determine an acceptable sensitivity, e.g. ``at least 50\% of known interactions should be recovered as very likely, having posterior probability of at least 75\%.'' 

We illustrate these considerations by analyzing the various models fit to the African Frugivory case study. For each model, the details of the specification, the pseudo-precision and recall of the sample, and the number of new interactions proposed are summarized in Table \ref{tab:SM_CVsummary}. The COIL 0/100 model appears to overpredict interactions with a very large number of new interactions predicted, very high recall, but poor discrimination via pseudo-precision. The proposed default prior COIL/COIL+ 75/100 yields very high pseudo-precision due to a lower overall prevalence but performs more poorly with respect to recall, failing to meet the sensitivity threshold proposed above. Additionally, the pseudo-precision levels for this model could be suspiciously high, being slightly above the upper bound of ten suggested above. The expert-defined scenarios yield the best balance of pseudo-precision and sensitivity and meet our rough prevalence and sensitivity guidelines proposed above. 

There is a substantial difference in the number of new interactions predicted between  models using the 75/100 and expert-defined probability scenarios - with far more new interactions predicted under the expert scenario. Recall from Tables \ref{tab:SM_postocc_frugs} and \ref{tab:SM_postoccs_plants} that the 75/100 prior results in relatively high posterior occurrence probabilities across the board; hence non-edges are more likely to be treated as observable and informative, pulling down interaction propensities. 

\begin{table}[ht]
\resizebox{\columnwidth}{!}{%
\begin{tabular}{@{}llllllrr@{}}
\toprule
Model & Prior & Occurrence Sampling   & $\frac{mean(\hat{p}_{cv})}{mean(\hat{p}_{all})}$  & $\hat{p}_{cv} > 0.5$ & $\hat{p}_{cv} > 0.75$  & \# New $\hat{p} > 0.5$  & \# New $\hat{p} > 0.75$  \\ \midrule
COIL 0/100 & 0/100\%    & NA                            &    1.85 &        0.92     &  0.81 &   104,588  & 66,100 \\
COIL 75/100 & 75/100\%  & Indicators only               &    12.45    &     0.65    & 0.48 &    3,221    & 1,059\\
COIL+ 75/100 & 75/100\%  & Indicators, Probabilities    &     11.42    &   0.66    & 0.48 &     4,315  & 1,509 \\
COIL Exp & Expert & Indicators only                     &    6.55    &    0.74     & 0.58 &     14,726 & 5,897 \\
COIL+ Exp & Expert & Indicators, Probabilities          &     7.17  &    0.78      & 0.57 &     12,822 & 5,637 \\ 
\bottomrule
\end{tabular}
}
\caption{Model summary: model descriptions, pseudo-precision (mean ratio), recall (at 50\% and 75\% thresholds), and the number of new interactions predicted at the  50\% and 75\% threshold. Pseudo and precision both refer to out of sample performance, while the number of new interactions are derived from full data model fits.}
\label{tab:SM_CVsummary}
\end{table}

\begin{figure}
    \centering
   \includegraphics[width=0.95\linewidth]{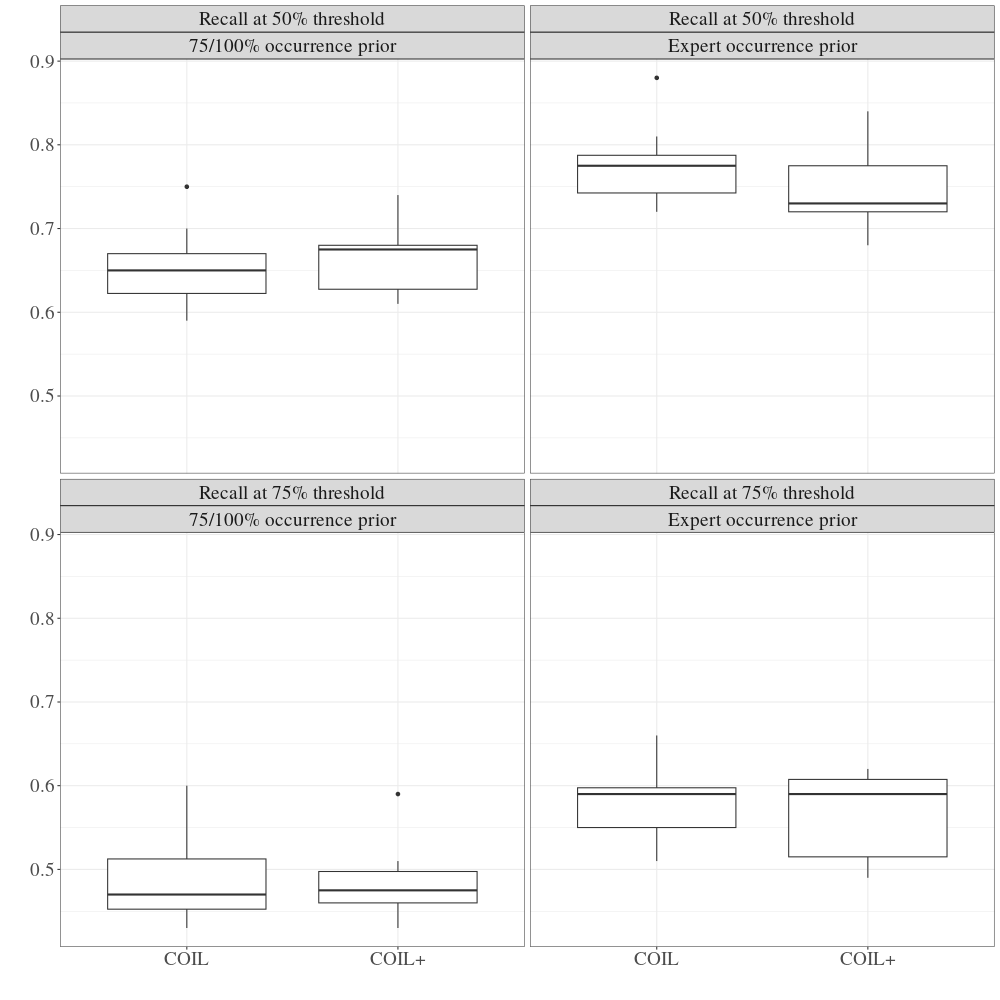}
    \caption{Out of sample recall performance as the proportion of true heldout interactions recovered with a posterior mean probability of at least 50\% and 75\% for two models under different occurrence priors.  Note that the original sampler using 0/100\% occurrence prior probabilities is omitted from this analysis to allow more clear visualization of the differences between the two best-performing models.}
    \label{fig:SM_recall}
\end{figure}

 To look more closely at recall for the two most plausible priors, see Figure \ref{fig:SM_recall}. We consider the original model sampling only the occurrence indicators (COIL) and the proposed model sampling both the occurrence indicators and the probabilities (COIL+), fit with the proposed default 75/100\% and the prior probabilities of occurrence defined by experts. Compared to the proposed default, the probability of events defined by the experts yields substantially higher recall at the 50\% and 75\% thresholds. Under the expert-defined prior, there are additional gains for COIL+ due to non-sampling of 0\% prior occurrence probability species under COIL. The COIL and COIL+ models have broadly similar recall performance using the 75/100 prior due to the absence of 0\% prior occurrence probability species in this construction.
 
\begin{figure}
    \centering
    \includegraphics[width=\linewidth]{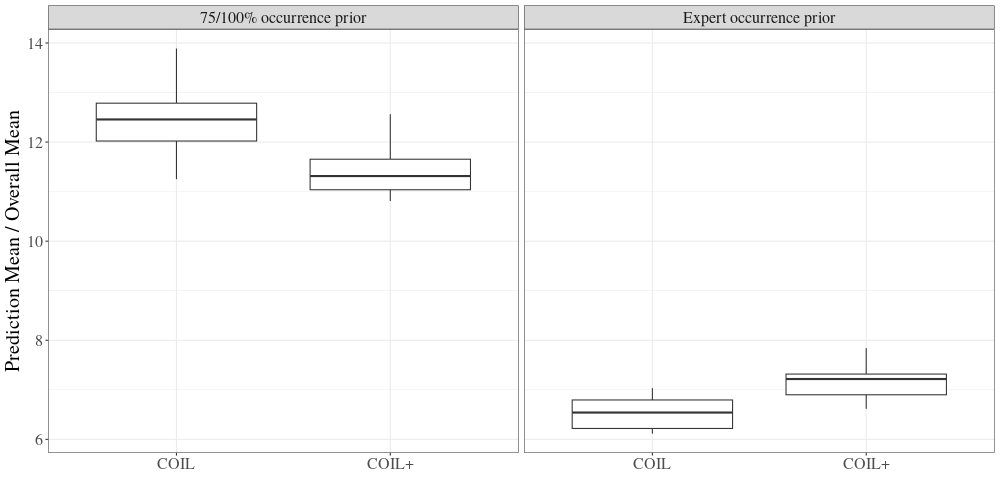}
    \caption{Out of sample pseudo-precision performance for  two models under different occurrence priors. We consider the original model sampling only occurrence indicators (COIL) and the proposed model sampling both occurrence indicators and probabilities (COIL+), fit with the proposed default 75/100\% and expert-defined occurrence prior probabilities. COIL 0/100 is omitted for more clear visualization of the relative performance of the best-performing models.}
    \label{fig:SM_precision}
\end{figure}

In terms of pseudo-precision, we see in Figure \ref{fig:SM_precision} that the expert-defined occurrence prior probabilities yield a slightly lower pseudo-precision, though still well above one; this corresponds to good discrimination of true interactions without sacrificing overall prevalence. Performance of COIL and COIL+ is broadly similar under the proposed default 75/100 with a more pronounced advantage to COIL+ under the prior defined by the expert, as was the case in the recall above. The strongly inferior performance of using 0/100\% occurrence prior probabilities is omitted from this analysis to allow for a more clear visualization of the relevant methods.

In summary, we find that the interaction-based occurrence construction of COIL 0/100 results in unreasonably high posterior interaction prevalence and comparatively low pseudo-precision. Compared to the expert-defined prior-occurrence probabilities, the proposed default prior results in a moderate decrease in out-of-sample recall and an increase in out-of-sample pseudo-precision which corresponds to a somewhat unexpectedly low posterior interaction prevalence. However, the loss of performance under the proposed default prior is minor compared to the dramatically worse interaction-based construction. Hence, in the event that no additional information about species occurrences can reasonably be provided, we recommend the default prior occurrence probability of 75\% for species not directly observed in an interaction at the site.

\subsection{MCMC Implementation and Diagnostics}
Posterior samples for the figures presented above and in the main text are obtained from four independent chains each consisting of 20,000 total iterations: 10,000 burn-in plus 10,000 iterations post burn-in thinned to 5\% resulting in 500 posterior samples. Note that while thinning is generally not recommended for Bayesian inference, in this case the high dimensionality of the posterior probability array entails high memory usage; thinning permits us to reduce the computational cost while running each chain for enough iterations to reach convergence. Because the model involves a large number of parameters, many non-identifiable, we obtain an overview of the sampler's convergence by monitoring the log-likelihood. We obtain a Gelman-Rubin statistic of 1.02 from post-burn-in samples, suggesting satisfactory convergence to the stationary distribution. 

Cross validation is performed using ten replicates, each with observed interactions for 100 distinct species pairs (frugivore $i$, plant $j$) heldout by setting the corresponding edges $A_{ijs} = 0$ and focus $F_{ijs} = 0$ for all studies $s$. Each cross validation replicate is implemented with a single MCMC chain 20,000 total iterations: 10,000 burn-in plus 10,000 iterations post burn-in thinned to 5\% resulting in 500 posterior samples (as for the full results described above). 

\subsection{Variable Importance}

The contribution of species covariates to the propensity for interaction is assessed by comparing a test statistic quantifying the association between observed covariates and interaction propensity with a permutation test statistic, as in \cite{PapadogeorgouJASA}. Specifically, let $X_{.l}$ be the vector of values for trait $l$ across all observed frugivores, and let $L_{.jr}$ be the vector of logit interaction probabilities for all frugivores and a given plant $j$ in posterior sample $r$. The observed test statistic is computed by averaging the squared correlation in these two quantities across all plants and all posterior samples: 

\begin{align*}
    \rho^2_{jlr} &= \text{corr}^2(X_{.l}, L_{.jr}) \\
    \hat{T}_l &= \frac{1}{n_P R} \sum_{j=1}^{n_P} \sum_{r = 1}^R \rho^2_{jlr}
\end{align*}

Permuted test statistics, each indicated by $T_{lb}^0$  for $b = 1, \cdots B$ permutations, is obtained by permuting the values of the covariate $X_{.l}$, and then performing a completely analogous computation to the above. Variable importance is then defined as the number of standard deviations by which the observed test statistic differs from the mean of the permutation test statistics. 

$$VarImp_l = \frac{| \hat{T}_l - \overline{T_l^o}|}{\text{sd}(T_l^0)}$$

The importance of the variable is defined as the number of standard deviations from the resampled mean the observed mean squared correlation is as in \cite{PapadogeorgouJASA}. The variable importance for the traits of the frugivore and the plant is plotted below in Figures \ref{fig:sm_frug_varimp} and \ref{fig:sm_plant_varimp}.

\begin{figure}
    \centering
    \includegraphics[width=0.75\linewidth]{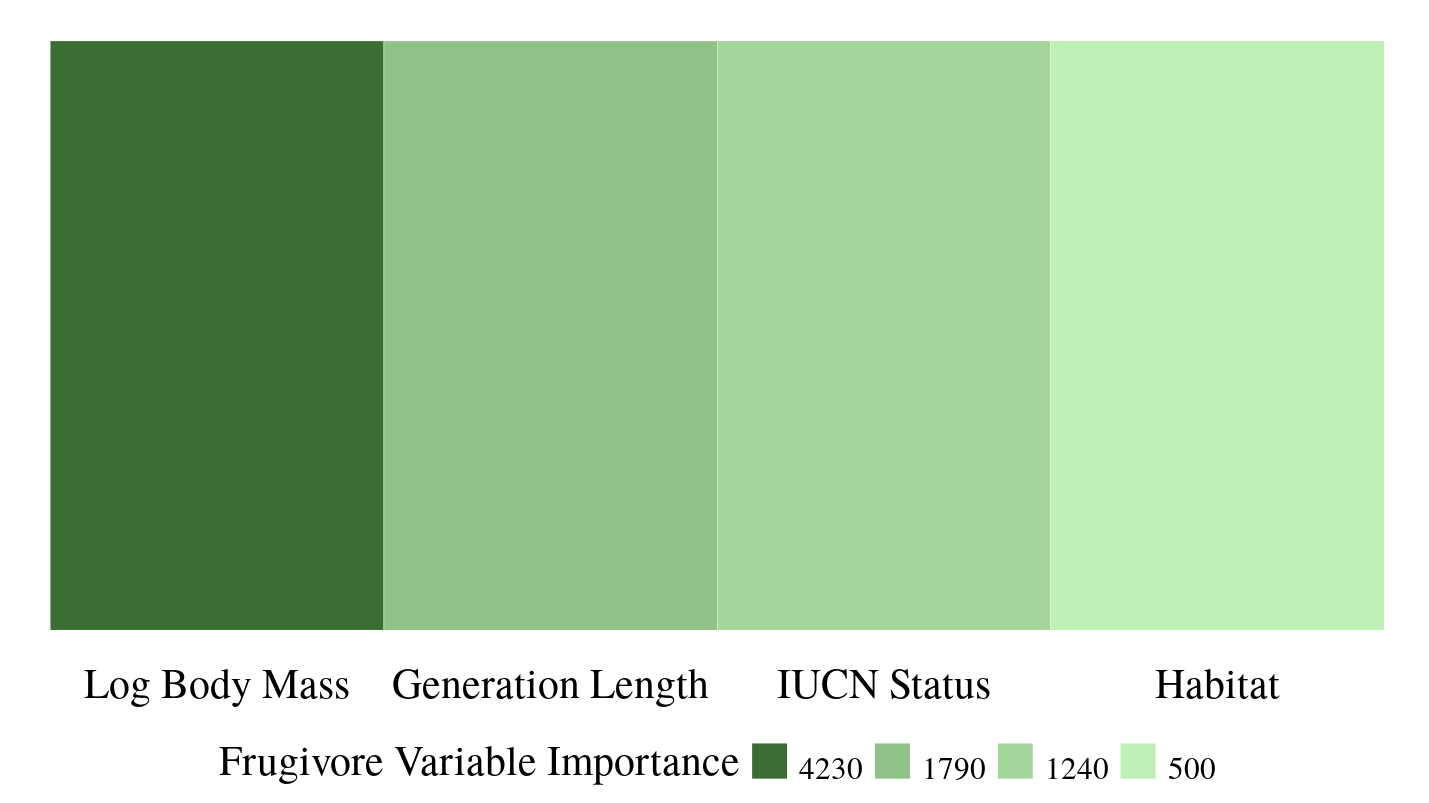}
    \caption{Frugivore covariates are plotted in order of decreasing importance from left to right. Color bars indicate the number of standard deviations away from the mean of the permuted test statistic that the observed test statistic is.}
    \label{fig:sm_frug_varimp}
\end{figure}

\begin{figure}
    \centering
    \includegraphics[width=0.7\linewidth]{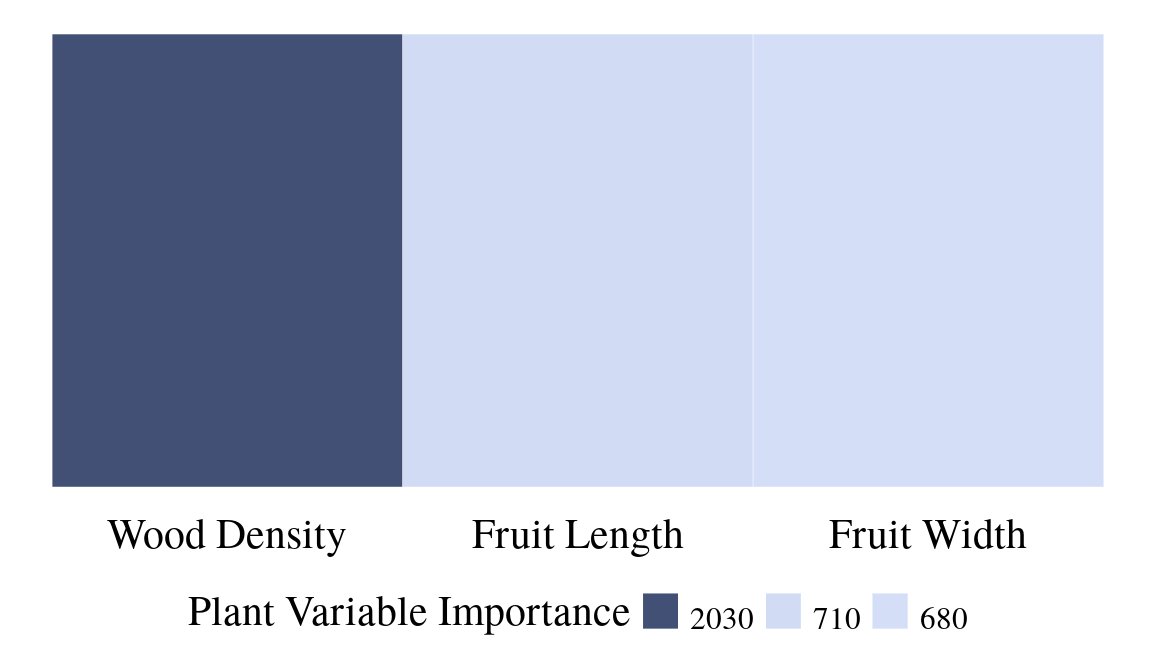}
    \caption{Plant covariates are plotted in order of decreasing importance from left to right.  Color bars indicate the number of standard deviations away from the mean of the permuted test statistic that the observed test statistic is.}
    \label{fig:sm_plant_varimp}
\end{figure}

\end{document}